\documentclass[10pt,conference]{IEEEtran}
\IEEEoverridecommandlockouts
\usepackage{cite}
\usepackage{amsmath,amssymb,amsfonts}
\usepackage{graphicx}
\usepackage{textcomp}
\usepackage{xcolor}
\usepackage{subfig}
\usepackage[font=footnotesize]{caption}

\usepackage{tikz}
\usetikzlibrary{arrows.meta,fit,positioning}

\usepackage[colorinlistoftodos,prependcaption,textsize=small]{todonotes}
\begin{document}

\newcommand{\circlenum}[1]{%
  \tikz[baseline=(char.base)]{
    \node[shape=circle,draw,line width=0.9pt,inner sep=1pt] (char) {\small{#1}};
  }%
}

\newif\ifdraft
%\drafttrue
\draftfalse

\ifdraft
\newcommand{\notel}[1]{\todo[inline,color=blue!40]{#1}}
\newcommand{\notem}[1]{\todo[inline,color=green!40]{#1}}
\newcommand{\noteh}[1]{\todo[inline,color=yellow!40]{#1}}
\newcommand{\notex}[1]{\todo[inline,color=red!40]{#1}}
\newcommand{\shihan}[1]{\todo[inline,color=purple!40]{#1}}
\else
\newcommand{\notel}[1]{}
\newcommand{\notem}[1]{}
\newcommand{\noteh}[1]{}
\newcommand{\notex}[1]{}
\newcommand{\shihan}[1]{}
\fi

\title{Incisor: Ex Ante Cloud Instance Selection\\ for HPC Jobs\\
% No acks needed for draft version
%\thanks{AA1, AA2 and AA3 were supported by an undisclosed grant. AA3 discloses financial interest in Anonymous Firm 1, which sells a commercial version of the Adviser software.}
}
%\thanks{S.C., M.L. and D.H. were supported by NSF Grant No. 2324735. D.H., discloses financial interest in Adviser Labs, Inc., which sells a commercial version of the Adviser software.}

\author{
\IEEEauthorblockN{Michael A. Laurenzano}
\IEEEauthorblockA{\small\textit{Institute for Software Integrated Systems} \\
\textit{Vanderbilt University}\\
Nashville, TN\\
michael.a.laurenzano@vanderbilt.edu}
\and
\IEEEauthorblockN{Shihan Cheng}
\IEEEauthorblockA{\small\textit{Institute for Software Integrated Systems} \\
\textit{Vanderbilt University}\\
Nashville, TN \\
shihan.cheng@vanderbilt.edu}
\and
\IEEEauthorblockN{David A. B. Hyde}
\IEEEauthorblockA{\small\textit{Institute for Software Integrated Systems} \\
\textit{Vanderbilt University}\\
Nashville, TN \\
david.hyde.1@vanderbilt.edu}
}

\maketitle

\begin{abstract}
We present Incisor, a cloud HPC job submission system for the \emph{ex ante instance selection problem}: choosing suitable hardware in the challenging but common setting where only the executable, inputs, and invocation commands are available at submission time. In practice, this task is manual and expertise-intensive, requiring users to combine incomplete knowledge of rapidly evolving cloud offerings with workload-specific intuition, static analysis, and systems reasoning to infer hardware constraints and select an instance type for each job. Incisor automates this process by pairing widely available program analysis tools with LLM-guided reasoning to infer hardware requirements and choose cloud instances. Using submission artifacts alone, Incisor atop frontier coding LLMs selects working AWS EC2 instances ex ante for 100\% of first-time runs of source-compiled (C, C++, Fortran) and Python applications. Against a strong baseline combining expert-derived constraints with SkyPilot’s instance selection, Incisor cuts job runtime by 54\% and instance costs by 44\%.

\end{abstract}

%\begin{IEEEkeywords}
%cloud computing, resource provisioning, instance selection, performance modeling
%\end{IEEEkeywords}

\section{Introduction}
\label{sec:introduction}

\begin{figure}[t]
\centering
\includegraphics[width=0.46\textwidth]{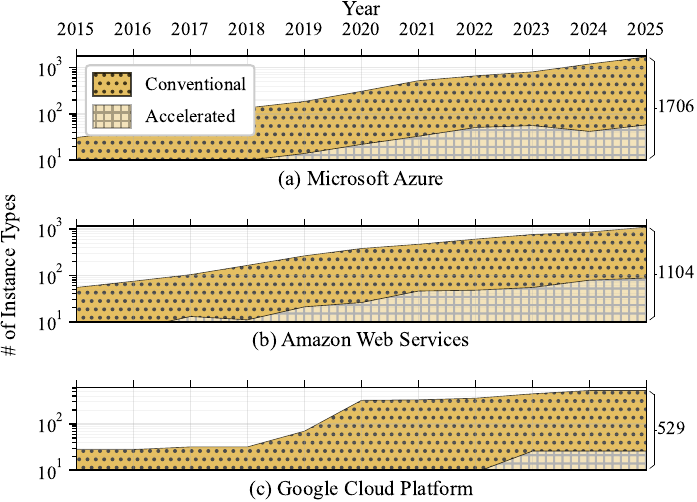}
\vspace{-1mm}
\caption{The number of available instance types from the major cloud providers has exploded over the past decade.}
\label{fig:instance-growth}
\end{figure}

In conventional HPC environments, hardware choices are amortized over long procurement cycles and typically result in a small number of stable choices. In the cloud, catalogs of hardware choices are large and evolve continuously. The major cloud providers expose thousands of resource configurations spanning wide ranges of architectures, CPU counts, memory capacities, accelerator options and storage configurations. Figure~\ref{fig:instance-growth} illustrates the growth of instance choices across the 3 largest public cloud providers: Microsoft Azure, Amazon Web Services and Google Cloud Platform. Today those 3 providers offer a combined choice of more than 3{,}000 instance types, an order of magnitude more choices than a decade ago.

% (.venv) mlaurenz@hyde06:~/adviser/instance_count_data/SC$ python3 ./plot_instance_counts_stacked.py

For users running HPC workloads in the cloud, the rapid growth of instance choices makes it increasingly important and increasingly difficult to address the \emph{ex ante instance selection problem}: choosing a well-suited hardware configuration for a job before the job is ever run. The user must navigate a large, expanding catalog under uncertainty about which hardware features the workload can exploit. In principle, making a good choice is straightforward for users with application domain knowledge, distributed systems expertise, and up-to-date knowledge of cloud offerings and prices. In practice, that combination of expertise is rare, and even when available, applying it to each workload is time-consuming, error-prone, and fragile to catalog or workflow changes.

Conventional approaches based on workload profiling and detailed performance analysis or prediction can be difficult to operationalize: they require instrumented runs, representative input datasets, multiple benchmarking iterations, and the resulting models must be updated as software versions, environments, and instance families change~\cite{kerbyson2001predictive, snavely2002framework, ipek2005approach, williams2009roofline, calotoiu2013using, tallent2014palm, venkataraman2016ernest, pham2017predicting, alipourfard2017cherrypick, pearce2023towards}. Moreover, established profiling and modeling techniques often rely on direct access to detailed hardware counters, topology details, and tracing tools, whereas cloud workloads typically run inside VM environments that occlude visibility into the underlying system's operation. For many users, these approaches add tooling overhead, domain expertise shortfalls, and extra cloud costs that outweigh the expected gains.

The hypothesis motivating this work is that agentic LLM systems are well matched to the ex ante instance selection problem. The key subproblems, inspecting job artifacts to estimate resource requirements and ranking hardware options against a cloud catalog, have four properties that favor an agentic approach. First, the evidence available to solve the problem is concrete and specific rather than subjective, consisting of executables, configuration variables, and cloud catalogs. Second, mature, widely available tools such as disassemblers, decompilers, and catalog APIs exist to extract that evidence. Third, the outputs are empirically verifiable by running the job and comparing inferred constraints to measured behavior. Finally, the problem tolerates imprecision because instance resources are discretized into coarse tiers (e.g., GBs of memory offered in powers of 2) where the risks of being wrong are asymmetric: underprovisioning may trigger failure while overprovisioning is merely inefficient.

\noteh{Space Saving: This paragraph in the intro is a bit duplicative of Section IV.A. I think we could cut it down a lot, maybe even to like one sentence, and just point to Section IV.A there. Not necessary, but, if we are needing to cut space (seems we're slightly over right now)}

In this paper we introduce \emph{Incisor}, an end-to-end cloud HPC job submission system that uses agentic LLMs designed to harness these properties to estimate hardware constraints and select cloud instances for HPC jobs given only the artifacts available before execution, including the program executable, its environment variables, inputs, and invocation commands. Incisor covers the lifecycle of submitted jobs, storing and indexing artifacts for reuse, collecting coarse-grained performance measurements during runs to refine future estimates, and triggering retries when instance recommendations prove unsuitable. This paper makes the following contributions:

\begin{itemize}

\item \textbf{Agentic Design for Ex Ante Instance Selection.} We characterize the properties that make the problems of ex ante hardware constraint estimation and instance selection amenable to agentic LLM automation, and describe the enabling agent architecture, tool suite, and verification mechanisms used to automate both stages (Section~\ref{sec:agentic}).

\item \textbf{End-to-End Job Handling.} We describe the design of Incisor, an end-to-end system for handling HPC job submissions given only the program executable, its inputs and the invocation commands. In addition to assigning jobs to instances to address the ex ante instance selection problem, Incisor collects and leverages high level performance measurements to improve instance selections on future runs; deals with erroneous predictions gracefully; and catalogs the artifacts necessary to minimize user burden in making cloud instance choices (Section~\ref{sec:system}).

\item \textbf{Evaluation Across 31 HPC Workloads.} Across 31 scientific, financial, and machine learning workloads spanning common parallelization mechanisms (OpenMP, MPI, Python multiprocessing), source languages (C, C++, Fortran, and Python), and accelerator use, we evaluate 6 LLM backends spanning frontier commercial and open-weight agents. We find that frontier coding agents achieve 100\% first-run success, reducing instance costs by 44\% and job runtime by 54\% relative to expert-derived configurations and improving to 45\% and 56\% when leveraging data from a single prior run (Section~\ref{sec:evaluation}).

\end{itemize}

\section{Motivating Example}
\label{sec:motivation}

Choosing a cloud instance before a job has ever been run can benefit cloud workflow and provisioning systems~\cite{cheng2026adviser, yang2023skypilot}, and requires answering a set of questions that are easy to pose but challenging to answer programmatically: \emph{How much memory will the job use?}, \emph{How many CPUs can it effectively exploit?}, and \emph{Will it benefit from accelerators or faster storage?} In practice, answering these questions involves expertise and domain knowledge about the application, its invocation, and the available cloud hardware options. To concretize this challenge, consider the OpenMP version of the NAS Parallel Benchmarks conjugate gradient (CG) application~\cite{bailey1991parallel}, class D, compiled for \texttt{x86\_64} and executed with 8 OpenMP threads.

\subsection{Measured Behavior}

Running the job across multiple cloud instances and measuring its behavior (see Section~\ref{subsec:runtime-measurement}) reveals peak resident memory usage of approximately $16.5$\,GB, full core utilization up to 8 CPUs with no benefit from additional cores, no GPU activity, and minimal I/O. These measurements give exactly the information needed for instance selection.

\subsection{Ex Ante Program Analysis}

While runtime measurement is informative, profiling runs incur time and cost, require expertise to interpret, and may need to be repeated after changes to the program or its invocation. The motivating question is whether similar conclusions can be reached ex ante from submission-time artifacts alone. For this example, the answer is yes. Key constraint dimensions can be recovered directly from the executable and invocation context before the job is ever launched.

\noindent \textbf{CPU Count.}
In many HPC codes, the degree of parallelism is determined by explicit configuration of recognizable programming constructs. In the CG example, the job run script sets \texttt{OMP\_NUM\_THREADS=8}. A static inspection of the executable reveals OpenMP constructs within the method \texttt{conj\_grad}, including calls to \texttt{omp\_num\_threads}. Together, this evidence indicates that the program will attempt to create 8 OpenMP threads and is therefore likely to benefit from 8 CPUs, but will not exploit additional CPUs unless the thread count is changed at invocation time.

\noindent \textbf{Platform and ISA.}
Static inspection also reveals compatibility constraints. Using \texttt{readelf} to inspect the program's ELF64 header, we can extract its \texttt{machine} fields to identify that the binary contains \texttt{x86\_64} machine code. Disassembly via \texttt{objdump} can then reveal specific SIMD idioms and registers. The presence of AVX instruction mnemonics and use of \texttt{ymm} registers indicates AVX instruction set usage. Conversely, the absence of both \texttt{zmm} registers and AVX-512 mnemonics suggests no requirement for AVX512-capable CPUs.

\noindent \textbf{Memory.}
Memory requirements can likewise be inferred from the executable. Following control flow from the program's entry point shows that execution unconditionally reaches the compiler-supplied allocation helper routine \texttt{alloc\_space}. This routine makes a handful of calls to \texttt{malloc}, and inspection of the routine reveals patterns such as the following, from which the allocation size can be derived directly:

{\footnotesize
\begin{verbatim}
// allocate 251MB of heap memory
bf 00 52 bc 0f          mov    $0xfbc5200,%edi
e8 1f c8 ff ff          call   10a0 <malloc@plt>
\end{verbatim}
}

Summing the sizes of the dominant data structures allocated in this fashion within \texttt{alloc\_space} shows that the program requires approximately 16\,GB of heap-allocated memory.

\noindent \textbf{I/O and Acceleration.}
Scanning the executable for parallel I/O APIs (MPI-IO, HDF5, ADIOS), GPU libraries (CUDA, ROCm), and offload-related code reveals no evidence of parallel I/O or accelerator usage, ruling out benefits from premium storage and GPU-enabled instances.

% python3 ./plot_instance_memory.py 18
\begin{figure}[t]
\centering
\includegraphics[width=0.46\textwidth]{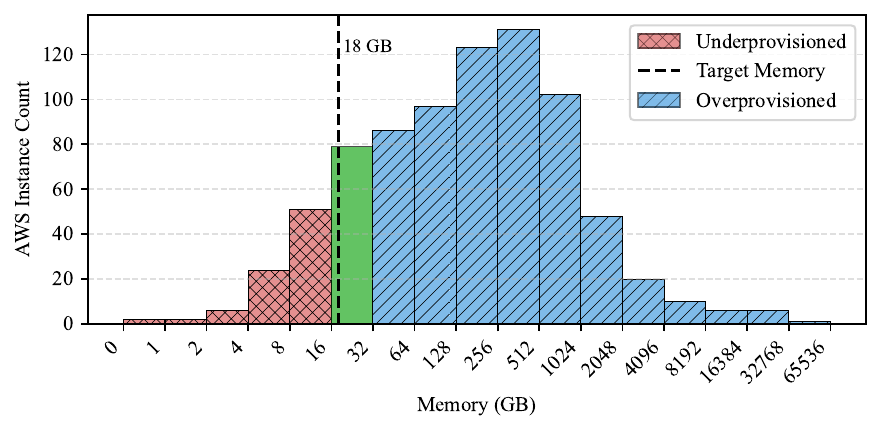}
\vspace{-2mm}
\caption{Memory capacity distribution across 794 \texttt{x86\_64} Linux AWS instance types. Instances below the 18\,GB threshold (red x pattern) for CG class D are expected to fail with an out-of-memory error.}
\label{fig:cg-motivation}
\end{figure}

\noindent \textbf{Instance Selection.}
Taken together, this evidence defines a submission-time constraint bundle for an \texttt{x86\_64} instance with at least 18\,GB of memory (16\,GB application heap plus overhead), 8 CPUs, AVX1 support, and no accelerators or premium storage. Figure~\ref{fig:cg-motivation} illustrates why this matters. Of 794 binary-compatible AWS instance types, 85 (11\%) fall below the 18\,GB threshold and would fail due to underprovisioning. Of the 709 feasible instances, only 79 fall in the right-sized 16--32\,GB tier, while the remaining 630 carry substantially more memory than needed, representing unnecessary cost from overprovisioning. Static analysis identifies this threshold without running the program, eliminating infeasible instances and focusing the search on cost-effective choices.

\subsection{The Opportunity for Automation}

The analysis above follows disciplined reasoning steps such as tracing control flow to allocation sites, inspecting the binary for ISA requirements, and cross-referencing findings against a cloud catalog. While no static analysis can ensure total coverage of program behavior, many HPC programs rely on standardized parallelization, allocation, and acceleration patterns whose resource implications are evident from submission-time artifacts. These properties makes the ex ante instance selection problem a natural fit for agentic LLM systems that can plan multi-step analyses, invoke tools, and programmatically derive actionable findings.

\section{Related Work}
\label{sec:related}

Prior work in LLM-assisted program analysis, cloud instance characterization and workflow scheduling informs Incisor's design. Each area addresses a facet of the problem, but none targets end-to-end instance selection for previously unseen HPC workloads using submission-time artifacts.

\noindent \textbf{LLM-Assisted Program Analysis.}
Recent work uses LLMs for program analysis tasks spanning static analysis, symbolic reasoning, and dataflow inference~\cite{li2023assisting, venkatesh2024emergence, chapman2024interleaving, wang2024llmdfa, wang2025contemporary}. A related line of work targets HPC performance analysis, including parallel program behavior prediction, performance modeling, and cost-aware cloud scheduling~\cite{bolet2025can, nichols2024hpc, cui2025llmperf, nguyen2024llmperf, pei2025llm}. Incisor differs by embedding agentic LLMs within an automated end-to-end system for handling job submissions via a constraint-driven instance-selection pipeline, emphasizing robustness to catalog changes and partial observability rather than standalone code reasoning.

\noindent \textbf{Cloud Instance Characterization and Selection.}
Other studies evaluate cloud instance families for HPC through empirical cost-performance analysis. Maas et al.\ explore instance options to identify cost-efficient choices, showing that the best selection is highly dependent on workload characteristics~\cite{maas2025exploring,maas2025investigating}. Kumar et al.\ compare GPU-enabled instances for HPC and AI workloads using a TOPSIS-based multi-criteria framework~\cite{kumar2024robust}, and Tharwani et al.\ show that CPU performance and cost efficiency vary substantially across instance generations and providers~\cite{tharwani2025evaluating}. Other work in this area automates configuration selection via exploratory runs. CherryPick~\cite{alipourfard2017cherrypick} uses Bayesian optimization and Ernest~\cite{venkataraman2016ernest} extrapolates from small-scale samples, but both require executing the workload. Incisor targets the challenging ex ante setting where no prior runs can be assumed to exist, inferring a suitable instance from submission-time artifacts alone while treating runtime history as helpful but optional. These empirical characterization studies are complementary to our approach, as their findings may inhabit LLM training corpora and be leveraged implicitly during agent reasoning about instance trade-offs.

\noindent \textbf{Resource Brokering and Workflow Scheduling.}
Prior work on mapping jobs to heterogeneous resources spans grid-era resource brokers such as AppLeS~\cite{berman2003adaptive} and Nimrod/G~\cite{bethwaite2010mixing}, which used performance models and economics-based scheduling to place jobs across distributed sites~\cite{krauter2002taxonomy}. Cloud workflow schedulers including Pegasus~\cite{pinto2020pegasus}, SkyPilot~\cite{yang2023skypilot}, and other DAG-based schedulers~\cite{juve2010scientific,rodriguez2014deadline} extended this to heterogeneous VMs, trading off cost, time, and reliability. These systems share the core problem of choosing where to run each task to optimize an objective such as cost-within-deadline. Incisor addresses a related but differently constrained problem, targeting per-job instance selection for previously unseen HPC workloads by inferring hardware requirements from submission-time artifacts alone and selecting from large, continuously evolving cloud catalogs.

\section{Agentic Instance Selection}
\label{sec:agentic}

\begin{figure*}[!t]
\centering
\includegraphics[width=0.90\textwidth]{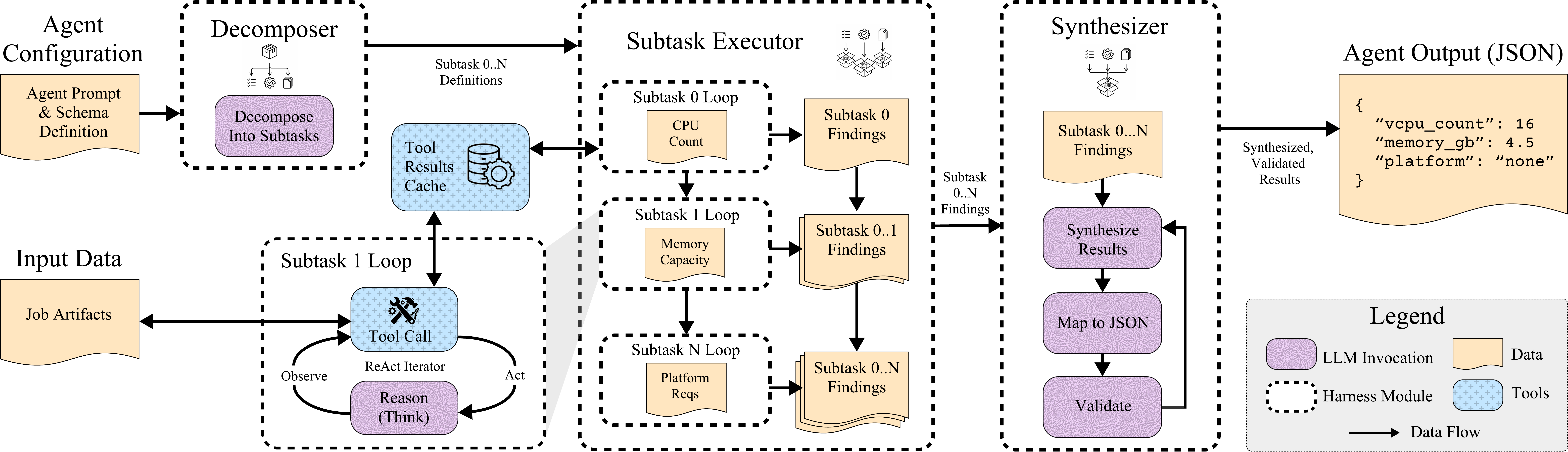}
\caption{Incisor agent architecture operation for hardware constraint estimation.}
\label{fig:agent-architecture}
\end{figure*}

The problem addressed in this work is choosing a cloud instance type for an HPC job given only the executable, environment variables, inputs, and invocation commands available at submission time. We decompose this into two subproblems:
\begin{itemize}
  \item \emph{Hardware constraint estimation}, which infers the job's resource requirements from submission artifacts alone, including memory high water mark, CPU count, platform and accelerator needs, and I/O intensity.
  \item \emph{Instance selection}, which maps those estimated constraints onto the catalog of available cloud instances to identify cost-effective choices.
\end{itemize}
We decompose the problem this way because the two subproblems draw on different evidence and tools. Constraint estimation draws on program analysis such as tracing allocations, identifying parallel constructs, and inspecting binaries, while instance selection draws on cloud offerings, pricing, and hardware capabilities. Crucially, this decomposition makes the intermediate constraint estimates independently checkable against measured resource use after the job runs, providing concrete diagnostics that can be incorporated into later runs.

\subsection{Why an Agentic Approach?}

The problems of submission-time constraint estimation and instance selection have a set of characteristics that make them particularly amenable to an agentic solution.

\noindent \textbf{Evidence-Grounded Reasoning.} Both subproblems admit to concrete evidence over subjective judgment, be it submission artifacts such as binaries, scripts, and configuration variables for constraint estimation or cloud instance catalogs for instance selection. Because the inputs to both stages are concrete and tool-accessible, the agent's reasoning can be grounded in factual evidence and derived observations at every step.

\noindent \textbf{Robust Tooling.} Mature tool interfaces exist for both stages, from disassemblers, decompilers, and symbol inspectors for constraint estimation to catalog query tools for instance selection. A tool-capable agent can gather evidence programmatically rather than relying on parametric knowledge alone.

\noindent \textbf{Verifiable Results.} Both subproblems produce outputs that can be empirically checked by running the job. Measured resource use validates inferred constraints and job success or failure validates the instance choice, creating a tight feedback loop that can quickly correct flawed reasoning across runs.

\noindent \textbf{Tolerance for Imprecision.} Cloud instance resources are quantized into coarse tiers or discrete options, with memory and CPU counts roughly doubling between adjacent sizes and features such as accelerators and fast storage either present or absent. This makes instance selection fundamentally a coarse classification problem rather than one requiring fine-grained quantitative estimates. The risks are asymmetric because underprovisioning causes hard failures while overprovisioning merely incurs inefficiency, favoring conservative estimates when evidence is incomplete. Unlike conventional HPC performance modeling, which is often useful only when predictions are precise, this problem is forgiving enough for an LLM-driven approach to be practical.

Concrete evidence, reliable tools, empirical verifiability, and tolerance for imprecision together map directly onto the strengths of modern agentic LLM systems~\cite{xie2023openagents, yang2023auto, chen2023autoagents, sahoo2024systematic, wu2024autogen, wolflein2025llm, mohammadi2025evaluation, googleadk2025}, which combine multi-step reasoning with structured tool use and evidence synthesis to automate sophisticated tasks.

\subsection{Agent Architecture}

Figure~\ref{fig:agent-architecture} shows the architecture of an Incisor agent, illustrated for the constraint estimation stage. Incisor supports both commercial and open-weight agent implementations. For open-weight models, Incisor uses a custom agent harness with vLLM for backend inference. For commercial models, the custom harness is replaced by the provider's agentic infrastructure. Implementation details are given in Section~\ref{sec:methodology}. Each agent receives an agent configuration that provides the driver prompt and output schema defining the task, and input data that provides the task-specific evidence. For constraint estimation, the input data comprises the submitted executable, invocation command, and environment variables.

Inspired by least-to-most prompting~\cite{zhou2022least}, the decomposer first invokes an LLM to break the configured task into a sequence of subtask definitions to produce findings that inform later subtasks. For the constraint estimation agent, these correspond to individual constraint dimensions such as CPU count, memory capacity, and platform requirements. The subtask executor then processes each subtask through a subtask loop driven by a ReAct iterator~\cite{yao2022react} to alternate between reasoning, acting by invoking tools, and observing tool results. The results of tool invocations are stored in a shared tool results cache so that results gathered for one subtask are available to later subtasks without redundant calls. Each subtask loop produces subtask findings that accumulate across the sequence, allowing later subtasks to build on earlier evidence.

\notem{``subtask'' appears too densely here}

Once all subtasks complete, a synthesizer accumulates findings, merges them into a coherent result, maps the result to the output schema as a structured JSON bundle, and validates the output against the schema constraints to produce the agent output. Incisor runs this architecture twice in sequence for each job. The constraint estimation agent output becomes input data to the instance selection agent, which additionally has access to the cloud instance catalog. The underlying agent harness enforces per-tool timeouts along with iteration and token budgets to ensure convergence instead of unbounded reasoning loops, while the prompting strategy biases toward conservative estimates when evidence is incomplete.

\subsection{Hardware Constraint Estimation}
\label{subsec:tools}

Upon job submission, Incisor triggers constraint estimation to turn submission-time artifacts into a structured constraint bundle for the subsequent instance selection stage.

\noindent \textbf{Input and Output.} The input data consists of the job submission artifacts: the executable or entry point, inputs, invocation command, environment variables, and optionally source code, documentation, or data from prior runs. The agent output is a hardware constraint bundle covering memory high water mark, CPU count, platform requirements (e.g., \texttt{aarch64} vs.\ \texttt{x86\_64}), GPU usage and count, GPU memory requirement, and I/O intensity. Each estimate includes a confidence level and rationale for use by later stages and offline inspection.

\noindent \textbf{Tools.} For executable binaries, the agent has access to standard Linux inspection tools including \texttt{file} and \texttt{readelf} for ELF metadata (ISA, architecture, shared library dependencies), \texttt{nm} for symbol tables, \texttt{ldd} for dynamic library resolution, \texttt{objdump} for disassembly, \texttt{size} for section sizes, and \texttt{strings} for embedded string literals. For deeper analysis, the agent can invoke \texttt{angr} for control flow graph construction and targeted symbolic exploration~\cite{wang2017angr}, and Ghidra for decompilation and cross-reference analysis~\cite{eagle2020ghidra}. For Python workloads, the agent reads the script and its local module tree directly. All file access, including any user-supplied auxiliary artifacts, is mediated through read-only directory listing, glob, read, slice, and grep tools scoped to staged artifact directories.

Prior job history is available via a job similarity tool built for Incisor. Executable similarity is assessed by constructing a static control flow graph from the submitted program using PyCG~\cite{salis2021pycg} for Python or \texttt{angr} for compiled binaries, which is then compared to historical job CFGs using the Weisfeiler-Lehman graph kernel~\cite{shervashidze2011weisfeiler} and subjected to a similarity threshold. When two executables match, a second comparison scores the invocation context (command line arguments and environment variables, cast to a typed graph) using the same kernel, and the top matching historical records are surfaced to the agent as additional evidence. Specifically, up to 3 recent successful runs and all unique failure cases with their failure reasons are returned, with recency used to break ties.

The agent also has access to a calculator tool for arithmetic operations and unit conversion, providing a reliable numerical capability to aid in translating low level program evidence into quantified intermediate or finalized observations.

\noindent \textbf{Verification.} Generated constraints are checked by running the job and comparing inferred resource needs against measurements, as described in Section~\ref{subsec:runtime-measurement}. Similarity-based history lookup allows these measurements and any prior failures to be used as strong evidence of constraints (or lack thereof) on future submissions of similar jobs.

\subsection{Instance Selection}

Running after constraint estimation, instance selection maps estimated constraints onto the catalog of available cloud instances to produce a ranked list of instance preferences.

\noindent \textbf{Input and Output.} The input data consists of the constraint bundle from the previous stage together with access to the cloud instance catalog. Incisor leverages SkyPilot's~\cite{yang2023skypilot} instance catalog, a provider-normalized store exposing a unified view of instance capabilities (memory in GB, vCPUs, physical CPUs, accelerators) and price points in dollars per hour, allowing clean comparisons to be made across providers. The agent output is a ranked list of up to 10 instance preferences with written rationale explaining each selection.

\noindent \textbf{Tools.} The agent queries the instance catalog to filter and compare offerings against hard constraints and reason about cost-performance trade-offs. The agent also has access to the same calculator tool used in constraint estimation.

\noindent \textbf{Verification.} Instance suitability is verified in two ways. At provision time, ranked candidates are validated against hard constraints, catching errors such as nonexistent instance types, instances with memory below the estimated requirement, or missing accelerator support before any cloud resources are allocated. At runtime, executing the job reveals conclusively whether the provisioned instance was adequate. Out-of-memory terminations, illegal instruction faults, and other resource-related failures indicate underprovisioning, while successful completion with low utilization indicates overprovisioning. Failures at either stage triggers redeployment to a more appropriate instance as described in Section~\ref{subsec:recovery}, and all outcomes serve as evidence to refine future estimates.

\section{Incisor Design and Implementation}
\label{sec:system}

\begin{figure*}[t]
\centerline{\includegraphics[width=0.90\textwidth]{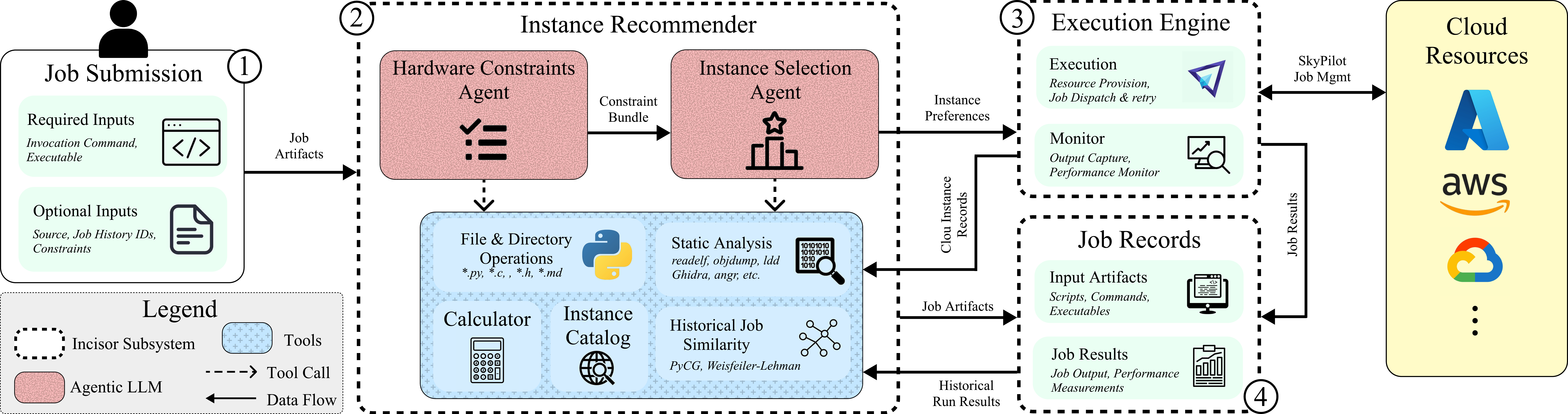}}
\caption{Overview of the Incisor system's job handling steps.}
\label{fig:incisor-overview}
\end{figure*}

Incisor embeds the agentic capability described above into an end-to-end job system that accepts job submissions, provisions cloud resources, executes jobs, and records measurements to feed back into future recommendations.

\subsection{System Overview}

Incisor begins with a user's submitted job artifacts, including the runnable workload together with its invocation context and any optional supporting artifacts. From those artifacts, it derives hardware constraints, ranks candidate instances from a cloud instance catalog, provisions resources, executes the job, captures measurements and outputs, and stores the resulting records for future reuse. Figure~\ref{fig:incisor-overview} depicts this workflow as four tightly integrated subsystems:

\begin{itemize}
  \item \circlenum{1} \textbf{Job Submission} ingests binaries, scripts, environment variables, source code, documentation, and optional prior-run references, and normalizes them into a structured job representation used by the rest of the system.
  \item \circlenum{2} \textbf{Instance Recommender} consumes the normalized job representation and runs the two-stage agent pipeline described in Section~\ref{sec:agentic} to produce a ranked list of instance preferences with brief explanations of the choices.
  \item \circlenum{3} \textbf{Execution Engine} provisions the preferred instance, launches the workload within a standardized execution envelope, and collects the measurements needed to validate recommendations and diagnose failures.
  \item \circlenum{4} \textbf{Job Records} persists job artifacts, selected resources, logs, outputs, and historical run results as a durable job record that supports provenance, debugging, and future recommendation runs.
\end{itemize}

\noindent This organization is key to Incisor's design. By embedding the recommender within a job service loop, runs automatically validate and reinforce the recommendations they were built on.

\subsection{Job Submission and Artifacts}
\label{subsec:submission-artifacts}

Incisor's Job Submission \circlenum{1} subsystem ingests the job artifacts provided in a user's job specification and normalizes them into a structured representation. At a minimum, a submission provides a runnable workload, its environment and its invocation command. Additional artifacts can supply more evidence for constraint estimation, and user-supplied constraints can override automated inference entirely.

\noindent \textbf{[Required] Job and Invocation Context.}
The invocation context comprises the run command and its arguments, relevant environment variables (e.g., \texttt{OMP\_NUM\_THREADS}), and any wrapper scripts. Small changes here can dominate runtime behavior, so Incisor treats invocation context as a first-class input to both static analysis and job history matching. Users may provide a compiled binary, a shell script, or an interpreted entry point. For binaries, Incisor infers platform and ISA constraints from ELF metadata and disassembly; for interpreted workloads, architecture selection is left to the selection agent.

{\footnotesize
\begin{verbatim}
> incisor run "OMP_NUM_THREADS=16 ./cg.D.x"
> incisor run "./train_model.py --epochs 10"
\end{verbatim}
}

\noindent \textbf{[Optional] Additional Artifacts.}
When available, Incisor can consume source code, build files, documentation, and problem-class descriptors as optional evidence. Incisor also searches its job history for similar jobs using the similarity tool described in Section~\ref{subsec:tools}, surfacing matching records to the constraint estimation agent. Users may also supply explicit job identifiers, hardware constraints, or bypass recommendation entirely by naming an instance type.

{\footnotesize
\begin{verbatim}
> incisor run "OMP_NUM_THREADS=16 ./cg.D.x" \
      --job-src "/path/to/NPB3.4-OMP"
> incisor run "OMP_NUM_THREADS=16 ./cg.D.x" \
      --job-history 18726,18843
> incisor run "OMP_NUM_THREADS=16 ./cg.D.x" --ram 32
> incisor run "OMP_NUM_THREADS=16 ./cg.D.x" \
      --cloud aws --instance-type c8a.4xlarge
\end{verbatim}
}

\subsection{Instance Preferences}
\label{subsec:instance-recommendations}

The Instance Recommender \circlenum{2} turns the normalized submission into a ranked list of instance preferences, typically in a few minutes per job submission and independent of job runtime. Internally, it uses the two-stage agent pipeline described in Section~\ref{sec:agentic}, first estimating hardware constraints from the submitted artifacts and any retrieved history, then ranking cloud instances using those constraints together with instance capabilities and prices from a managed catalog.

Incisor's ranking procedure is designed to be deliberately conservative. Memory capacity, architecture and ISA compatibility are treated as hard feasibility constraints because underprovisioning can easily lead to job failure. Other dimensions, such as CPU count above the application's parallelism ceiling, premium storage, or accelerators when unused, are treated as softer performance and cost tradeoffs. When uncertainty remains after submission-time analysis, such as when memory bounds are inferred from partial evidence, the recommender biases toward instances with headroom and records that uncertainty in the explanation. As part of this conservative posture, the constraint estimator enforces a 1\,GB floor on memory estimates and the instance selector enforces a 2\,GB floor when mapping constraints to instances, ensuring headroom for the operating system, monitoring facilities, and other non-application overhead, even for minimal workloads.

In the absence of strong evidence to the contrary, the recommendation pipeline also leans toward newer CPU families, which often offer better price-performance, and toward AMD over Intel architectures based on observed trends in common cloud catalogs~\cite{maas2025exploring, maas2025investigating}. These heuristics act only as a disposition and are overridden by stronger signals from inferred constraints and job history.

\subsection{Resource Provisioning and Execution}

Incisor's Execution Engine \circlenum{3} turns instance preferences into an actual job run. It provisions cloud resources and executes workloads using SkyPilot~\cite{yang2023skypilot}, which provides cloud-agnostic job orchestration, access to provider catalogs, and portable provisioning across clouds. Incisor uses SkyPilot's provisioning substrate but replaces its built-in instance selection with the recommender described above, then layers artifact storage, indexing, and job-history management on top to support richer reasoning across runs. The instance preferences provide both the default instance choice for execution and fallback candidates when the top-ranked choice is unavailable or proves to be a poor fit at runtime. When the preferred instance cannot be provisioned or fails to complete execution, Incisor steps down the list until a working candidate is found.

Jobs are launched under a standardized execution envelope with consistent lifecycle semantics covering setup, run, and teardown phases, uniform log capture, and a predictable directory layout for staged artifacts and outputs. This standardization is important for the rest of the system because it makes monitoring facilities portable across instance types and cloud providers and ensures that failures and outputs are recorded in a form that can be reused by the job record and feedback mechanisms described below.

\subsubsection{Runtime Performance Monitoring}
\label{subsec:runtime-measurement}

To make recommendation outcomes checkable, Incisor collects runtime measurements during execution across heterogeneous cloud instances and providers. Incisor deploys a uniform monitoring stack based on Prometheus, a widely adopted open source monitoring system with broad cloud support~\cite{turnbull2018monitoring}. Our design prioritizes measurements that are widely available across Linux distributions and cloud images, stable under common containerization and orchestration choices, and directly interpretable in terms of the constraint dimensions targeted by the recommender. Table~\ref{tab:runtime-metrics} summarizes a selection of key measurements made, their sources, and how Incisor uses them.

\begin{table}[t]
\centering
{\setlength{\tabcolsep}{1.3pt}
\begin{tabular}{@{}p{0.22\linewidth}|p{0.30\linewidth}|p{0.46\linewidth}@{}}
\hline
\textbf{Purpose} & \textbf{Metrics} & \textbf{Source} \\
\hline
\hline
Mem. Capacity & \texttt{MemTotal}, \texttt{MemAvailable} & \texttt{/proc/meminfo} (D) \\
\hline
Mem. Stalls & \texttt{some}, \texttt{full} & \texttt{/proc/pressure/memory} (D) \\
\hline
CPU Utilization & \texttt{user}, \texttt{system}, \texttt{idle} & \texttt{/proc/stat} (D) \\
\hline
CPU Utilization & \texttt{waiting}, \texttt{running} & \texttt{/proc/schedstat} (D) \\
\hline
CPU Stalls & \texttt{some}, \texttt{full} & \texttt{/proc/pressure/cpu} (D) \\
\hline
GPU Compute & \texttt{GPU-Util} & \texttt{nvidia-smi} (V) \\
\hline
GPU Memory & \texttt{Memory-Usage} & \texttt{nvidia-smi} (V) \\
\hline
I/O Activity & \texttt{read}, \texttt{write} & \texttt{disk\_bytes} (D) \\
\hline
I/O Stalls & \texttt{some}, \texttt{full} & \texttt{/proc/pressure/io} (D) \\
\hline
Disk Space & \texttt{Used}, \texttt{Avail} & Linux utility \texttt{df} \\
\hline
\end{tabular}}
\caption{A selection of key runtime metrics collected by Incisor. Most are gathered via Prometheus using its node exporter (D)\cite{nodeexporter2026} and NVIDIA GPU exporter (V)\cite{gpuexporter2026}.}
\label{tab:runtime-metrics}
\end{table}

Each measurement category maps directly to a constraint dimension used by the recommender. Memory capacity and pressure-stall indicators validate the memory high water mark estimate, CPU utilization and scheduling statistics validate CPU count, GPU utilization and memory confirm accelerator sizing, I/O throughput and pressure surface storage bottlenecks, and disk space measurements inform disk provisioning. Together these measurements support assessment of whether the selected instance was well-matched to the workload.

For Python workloads, module dependencies are initially gathered ex ante by the constraint estimation agent from source code analysis, since the true set of imported modules is only known after execution. To close this gap, Incisor deploys a lightweight runtime harness that records the full set of modules imported during execution, providing the constraint estimation agent with concrete module dependencies on future runs.

\subsubsection{Recovery from Underprovisioning}
\label{subsec:recovery}

Underprovisioning resources may lead to failures such as out-of-memory errors, illegal instruction faults, missing GPUs, or insufficient disk. When such a failure occurs, Incisor retries the job on the next ranked instance to address the diagnosed violation. For example, an out-of-memory failure triggers a retry on the next candidate with at least 2$\times$ the memory. This is possible because the recommender produces a ranked candidate set rather than a single choice, and because many such errors are recoverable after one observed run.

\subsection{Job Results}

Job Records \circlenum{4} persists job results and provenance as first-class system outputs. For each completed run, Incisor stores job artifacts, the selected instance configuration, logs, exit code, outputs, and raw performance measurements. From these, Incisor derives and stores a compact summary of the measurements described in Section~\ref{subsec:runtime-measurement}, small enough ($<$\,1k tokens) to avoid straining LLM context windows. This allows multiple historical run results to be surfaced to the constraint estimation agent on future runs of similar jobs, closing the loop between inference and execution.

%\section{Methodology}

\section{Results}
\label{sec:evaluation}

We describe our experimental setup and evaluate whether Incisor can reliably select viable, cost-effective instances from submission-time artifacts alone. We also investigate the factors that drive result quality, including the richness of available inputs and the capability of the underlying LLM.

\subsection{Experimental Setup}

\label{sec:methodology}

\noindent \textbf{Infrastructure.} Incisor supports both commercial and locally-hosted LLM agents. All agents use identical prompt templates, I/O schemas and verification logic. We evaluate two commercial models, Claude Opus~4.6 and Claude Sonnet~4.6, both using the Claude Code~2.1 harness for task execution, tool orchestration, and structured output parsing. Where the production harness provides built-in tooling capabilities, Incisor leverages those tools in place of our custom tool implementations. For open-weight models, we use Qwen3.5 35B A3B, Nemotron Super 49B, Llama3.3 70B, and Ministral3 14B via the agentic harness described in Section~\ref{sec:agentic}. This agentic harness is written in Python, with backend inference performed via vLLM~0.16~\cite{kwon2023efficient} on a control plane server equipped with a 64 core AMD EPYC 9575F and 4 NVIDIA RTX 6000 Ada GPUs (48\,GB each) to enable parallel open-weight model inference. Commercial backends require no dedicated local LLM inference hardware.

Resource provisioning and execution use SkyPilot~0.12.0 along with a Prometheus 3.9 monitoring stack detailed in Section~\ref{subsec:runtime-measurement}, both of which operate uniformly across cloud providers. We evaluate Incisor on Amazon Web Services (AWS) within a single region (\texttt{us-east-1}) using on-demand EC2 instances and prices frozen to a single point in time across all runs, ensuring that results reflect constraint estimate and instance selection quality rather than regional availability factors and/or price fluctuations across time or providers.

\noindent \textbf{Baselines.} We compare Incisor against two baselines. The first, \emph{Expert + SkyPilot}, represents a strong baseline in which an expert manually specifies key hardware constraints for each workload. The expert constraints were derived by a researcher with over 20 years experience in HPC/systems performance analysis via a combination of static analysis of program artifacts and iterative refinement over multiple job runs. These constraints are passed into SkyPilot via its \texttt{resources} block, then SkyPilot selects the cheapest feasible instance satisfying those resource constraints. This baseline represents a realistic outcome of the manual, iterative approach outlined in Section~\ref{sec:introduction}. It reliably produces viable instance selections but requires significant per-workload effort from a domain expert that remains difficult to scale to large numbers of workloads and to users lacking deep HPC experience. The second baseline, \emph{SkyPilot}, uses SkyPilot default instance selection with no user-supplied hardware constraints beyond the run command, representing what a user would experience when using a state-of-the-art cloud provisioning tool but bringing no particular workload or cloud expertise to provide useful guidance on instance selection.

\noindent \textbf{Workloads.} We use 31 benchmark configurations spanning HPC microbenchmarks, scientific simulations, computational finance, and machine learning, covering diverse parallelization and acceleration strategies and operating across compute bound, memory bound, and I/O bound resource regimes. For consistency with the SkyPilot baselines, which do not support automatic platform selection, all compiled workloads use GCC 12.2 to target \texttt{x86\_64}. Interpreted workloads use Python 3.11 and are architecture agnostic unless native extensions impose additional constraints derived by the instance recommender.

Microbenchmarks include \texttt{hpcg}\cite{heroux2013hpcg} (sparse linear algebra with irregular memory access) and \texttt{iozone}\cite{norcott2003iozone} (file I/O). From HeCBench \cite{jin2023benchmark} we use \texttt{bezier}, \texttt{invk2j}, and \texttt{nms}. From the NAS Parallel Benchmarks\cite{bailey1991parallel} we use the MPI and OpenMP variants of \texttt{bt}, \texttt{cg}, \texttt{ep}, \texttt{ft}, \texttt{lu}, \texttt{mg}, and \texttt{sp}, plus the OpenMP version of \texttt{ua}. Additional scientific applications include \texttt{lammps}\cite{thompson2022lammps} (molecular dynamics), \texttt{lulesh}\cite{karlin2013lulesh} (shock hydrodynamics), and two Palabos lattice-Boltzmann cases\cite{latt2021palabos}: \texttt{boussinesq3d} (Rayleigh--B\'enard convection) and \texttt{bubbles3d} (two-bubble collision). We use four Python applications, \texttt{mnist} (image classification), \texttt{astro} ($n$-body simulation), \texttt{genomics} (association analysis), and \texttt{seismic} (wave propagation). Financial workloads \texttt{blackscholes}, \texttt{bonds}, and \texttt{repo} are from FinanceBench\cite{grauer2013accelerating}.

% ./plot_cost_num1.py
\begin{figure}[!t]
\centering
\includegraphics[width=0.46\textwidth]{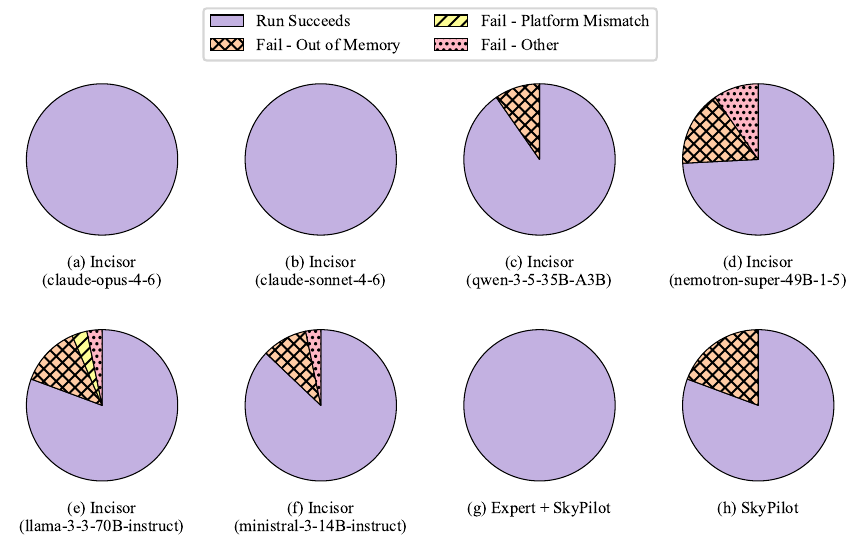}
\vspace{-1mm}
\caption{Job success rates using Incisor agent selections and SkyPilot baseline ex ante instance assignments.}
\label{fig:num1-failures}
\end{figure}

\begin{figure*}[!t]
% python3 ./plot_categorical_bars.py hierarchical-L0cmd-claudeopus hierarchical-L4bin-claudeopus hierarchical-L6hist-claudeopus
  \subfloat[All constraint estimates.\label{fig:categorical-predictions}]{%
    \includegraphics[width=0.234\textwidth]{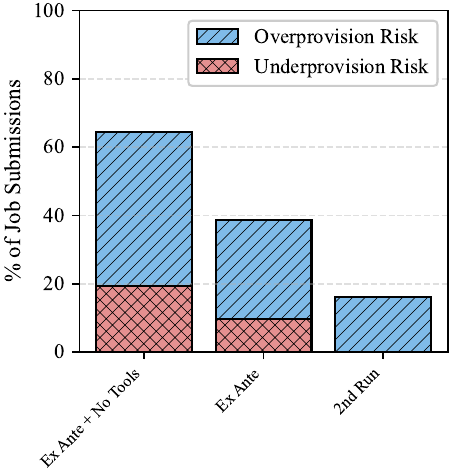}
  }
  \hfill
% python3 ./plot_memory_bars.py hierarchical-L0cmd-claudeopus hierarchical-L4bin-claudeopus hierarchical-L6hist-claudeopus
  \subfloat[Memory high water mark estimates.\label{fig:memory-predictions}]{%
    \includegraphics[width=0.74\textwidth]{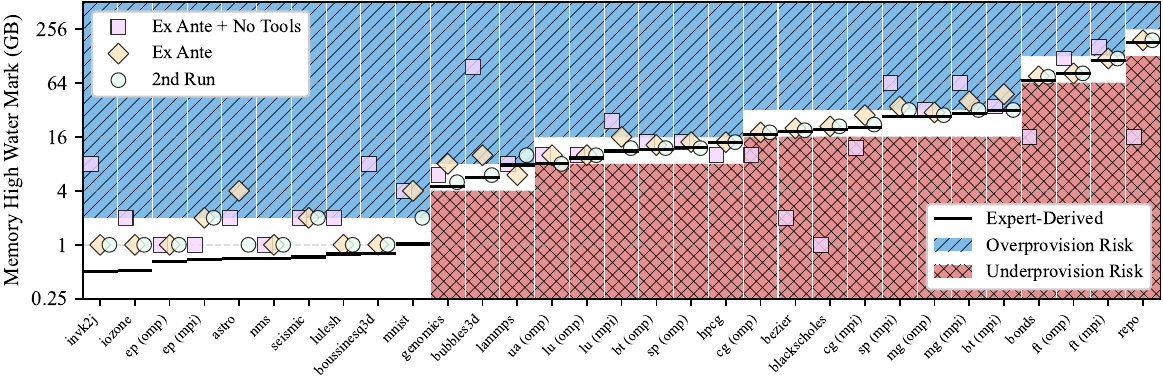}
  }
  \caption{Incisor constraint estimates across input configurations. (a) Fraction of job submissions with over- or under-provisioning risk, showing that tool access and historical data substantially reduce misprediction. (b) Per-application memory high water mark estimates under three input configurations compared to expert-derived ground truth (black lines). Blue (hatched) regions indicate overprovision risk; red (crosshatched) regions indicate underprovision risk.}
  \label{fig:constraint-estimates}
\end{figure*}

\subsection{Can an agentic LLM select instances that actually
run a submitted job on the first attempt?}
\label{subsec:first-run-success}

Figure~\ref{fig:num1-failures} illustrates success among instances chosen ex ante by Incisor using six LLM backends and by both the \emph{SkyPilot + Expert} and \emph{SkyPilot} baselines. Incisor with frontier coding models Opus4.6 and Sonnet4.6 achieve a 100\% first-run success rate across the 31 workloads, as does the Expert + SkyPilot baseline. By contrast, SkyPilot without expert-chosen constraints fails on 19\% of jobs, predominantly from out-of-memory errors, underscoring the importance of making use of hardware constraints for instance selection. Among open-weight models, results are more varied. Qwen3.5 fails on 10\% of jobs, all from out-of-memory errors. Nemotron has the highest failure rate at 26\%, followed by Llama3.3 at 20\% and Ministral3 at 13\%. Out-of-memory is the class of error that dominates across all configurations, indicating that memory estimation is the most relevant (and challenging) constraint to correctly provide in this domain. 

However, other error types arise in a handful of cases. Llama3.3 recommended the platform-incompatible instance \texttt{c6a.32xlarge} for \texttt{bonds}, choosing an instance lacking AVX-512 support. The associated job failed within 1 second of post-setup execution and was successfully rerun on the 2nd-ranked instance. Nemotron twice hallucinated a nonexistent instance \texttt{m8a.32xlarge}; while similar instances such as \texttt{m6a.32xlarge} and \texttt{m8a.24xlarge} exist, the requested one does not. This was caught at provisioning time and retried successfully. All \emph{Fail - Other} cases were recovered via Incisor's ranked-candidate fallback mechanism (Section~\ref{subsec:recovery}).

These results show that ex ante reasoning performed by the strongest models is accurate enough for practical deployment. To understand what makes it accurate, it is important to isolate the contribution of each class of submission-time input.

\subsection{What drives quality in agentic constraint estimates?}
\label{subsec:constraint-estimation}

Figure~\ref{fig:constraint-estimates} compares 3 configurations using Incisor + Opus4.6 with data sources of increasing richness. In \emph{Ex Ante + No Tools} the agent receives only the invocation command and environment variables, with no access to the executable or analysis tools. This configuration isolates reasoning from invocation context alone. In \emph{Ex Ante} the agent has the executable and access to the binary analysis tools described in Section~\ref{subsec:tools} and to source code for Python workloads. This setting reflects Incisor's primary operating mode. Finally, in \emph{2nd Run} the agent receives coarse-grained runtime measurements from one prior successful run, in addition to the executable and tools.

\noindent \textbf{All Constraint Estimates.} Figure~\ref{fig:constraint-estimates}(a) summarizes the fraction of workloads whose constraint estimates carry over- or underprovision risk across the 3 configurations. We classify estimates as underprovision risks when they underpredict a quantitative constraint or mispredict a categorical one in a way that would cause failure (e.g., incompatible platform ISA), and overprovision risks when they exceed requirements or choose more capability than necessary. The consequences are asymmetric, as underprovisioning can cause outright failure while overprovisioning wastes resources but the job succeeds. 61\% of submissions are mispredicted in some fashion under \emph{Ex Ante + No Tools}, with a substantial fraction risking underprovisioning. In this setting, it is trivial to confound the constraint estimator's CPU count estimates, for example by providing \texttt{OMP\_NUM\_THREADS=1024} when the job makes no use of OpenMP. Allowing for analysis of submission artifacts in \emph{Ex Ante} reduces misprediction to 36\% and, critically, cuts the underprovision risk rate in half.

% python3 plot_normalized_costs.py --plot-skypilot skypilot_cpumem.txt --label-ranked --ranked-limit 5 hierarchical-L4bin-claudeopus
\begin{figure}[t]
\centering
\includegraphics[width=0.49\textwidth]{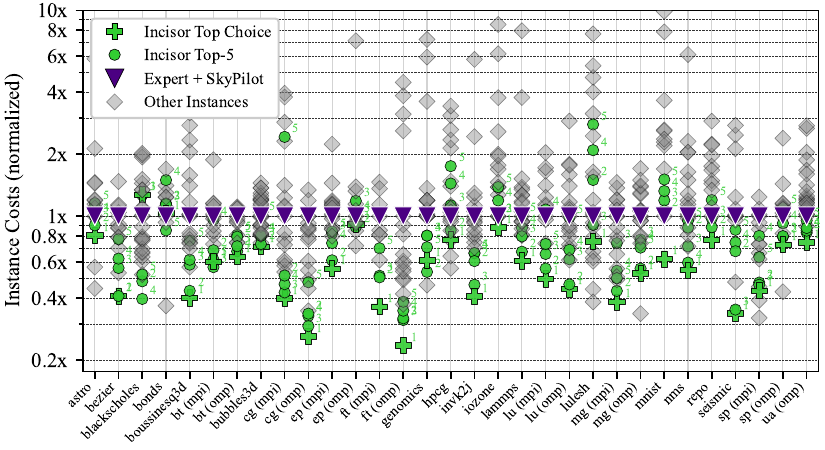}
\vspace{-6mm}
\caption{Instance costs for the top 5 Incisor-chosen instances compared to the \emph{Expert + SkyPilot} approach. Incisor's top choice (green +) achieves far lower instance costs than \emph{Expert + SkyPilot} (purple triangle) on most jobs.}
\label{fig:cost-efficiency}
\end{figure}

\noindent \textbf{Memory Estimates.} Figure~\ref{fig:constraint-estimates}(b) shows per-application memory high water mark estimates compared against expert-derived ground truth. The shaded regions indicate overprovision risk (blue, above ground truth) and underprovision risk (red, below ground truth), respectively. Under \emph{Ex Ante + No Tools}, the agent frequently defaults to brittle heuristics, in some cases producing estimates that are an order of magnitude above or below the true value. Under \emph{Ex Ante}, where the agent can access binary analysis tools for compiled applications and the source code of Python applications, estimates tighten substantially toward ground truth. The visible 1\,GB estimate floor on the left side of the figure reflects the conservative memory minimums enforced by the constraint estimator and instance recommender (see Section~\ref{subsec:instance-recommendations}).

\noindent \textbf{Runtime History.} With one prior job run (\emph{2nd Run}), misprediction falls from 36\% to 15\% (Figure~\ref{fig:constraint-estimates}(a)), with all risk accruing to the safer direction of overprovisioning rather than more dangerous underprovisioning. Once Incisor's job record feedback loop surfaces observed runtime behavior alongside new submissions, its future recommendations become both more accurate and more conservative. This shift toward safe overprovisioning reflects the design of Incisor's prompting and ranking logic, which prefer recoverable overestimation over failure-inducing underestimation.

These results show that invocation context alone is insufficient for reliable constraint estimation. Tool-based analysis of submission artifacts is the key enabler, and a single prior run further tightens estimates toward safe overprovisioning. Our initial experiments in supplying source code alongside compiled binaries (not shown) did not yield measurable improvements, suggesting that binary analysis tools already extract the most relevant signals for compiled workloads. This is plausible because the key constraint indicators like allocation sizes, thread counts, and ISA features, are visible in the compiled binary and directly accessible to analysis tools. Source code may offer greater benefit for more complex codebases or when compile-time optimizations obscure program structure, though further experiments in this direction are left as future work.

\subsection{How well do automatically chosen instances fit the job?}
\label{subsec:cost-runtime}

% python3 plot_normalized_costs.py --runtime --plot-skypilot skypilot_cpumem.txt --label-ranked --ranked-limit 5 hierarchical-L4bin-claudeopus
\begin{figure}[t]
\centering
\includegraphics[width=0.49\textwidth]{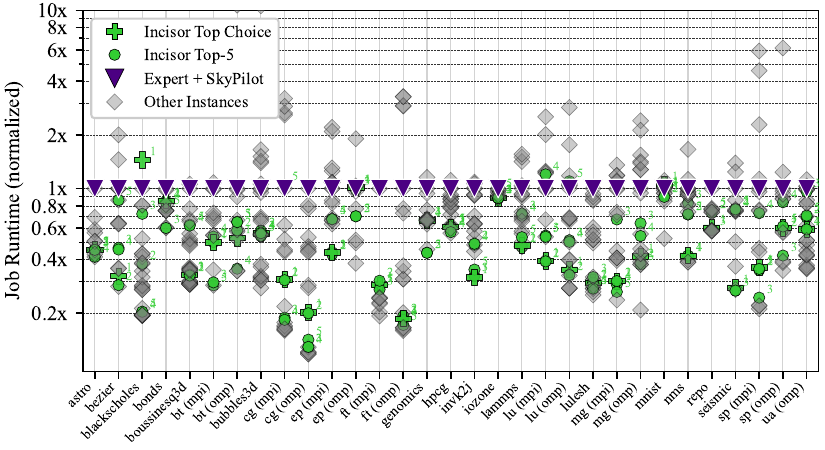}
\vspace{-6mm}
\caption{Runtime for the top 5 Incisor-selected instances normalized to the \emph{Expert + SkyPilot} approach. Incisor frequently selects instances that are substantially faster in execution time.}
\label{fig:runtime-efficiency}
\end{figure}

Figures~\ref{fig:cost-efficiency} and~\ref{fig:runtime-efficiency} show per-application instance costs and runtime normalized to \emph{Expert + SkyPilot} using Incisor + Opus4.6 with submission artifacts only and no prior runtime history, plotting the top-ranked instance (green +), remaining top-5 choices (green circle), the \emph{Expert + SkyPilot} baseline (purple triangle), and other feasible instances run during testing (gray diamond) for context.

\noindent \textbf{Instance Costs.} Across the full workload suite, Incisor's top-ranked instance reduces geometric mean instance costs by 44\% relative to \emph{Expert + SkyPilot} (Figure~\ref{fig:cost-efficiency}). These improvements arise because Incisor jointly reasons about feasibility and cost-performance tradeoffs, whereas \emph{Expert + SkyPilot} simply selects the cheapest instance satisfying expert constraints without reasoning about whether different feasible instances would be more cost-effective. Incisor's ranked preference list also provides fallback options, with multiple alternatives achieving costs near the baseline in most workloads. The few cases where Incisor's top choice costs more than \emph{Expert + SkyPilot} involve overestimated resource requirements.

\noindent \textbf{Job Runtime.} Incisor's top-ranked instances achieve 54\% lower geometric mean runtime than \emph{Expert + SkyPilot} (Figure~\ref{fig:runtime-efficiency}), often selecting instances that are both cheaper and faster. This is possible because the baseline optimizes hourly price, not total cost, while Incisor weighs feasibility, performance, and cost, steering toward newer-generation instance families with better price-performance ratios. These results demonstrate that coarse but grounded reasoning is sufficient to place workloads in the right region of the instance space without requiring precise numerical prediction.

\subsection{How does choice of the underlying model affect results?}
\label{subsec:model-comparison}

\begin{figure}[!t]
\centering
\includegraphics[width=0.48\textwidth]{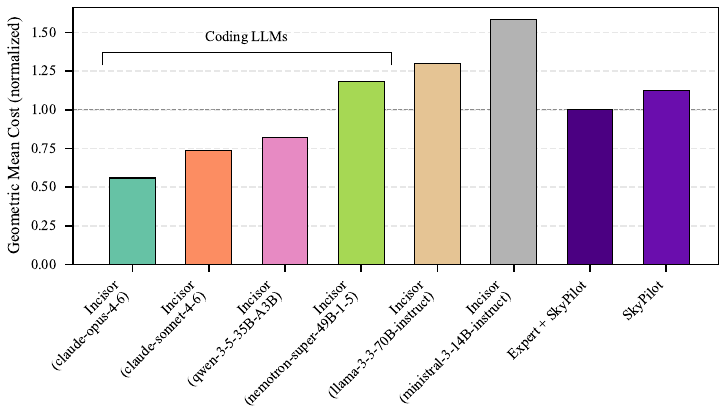}
\vspace{-3mm}
\caption{Geometric mean instance costs across Incisor and baselines, normalized to \emph{Expert + SkyPilot}. The top three coding-specialized LLMs outperform the expert baseline.}
\label{fig:num1-geomean}
\end{figure}

Figure~\ref{fig:num1-geomean} compares geometric mean instance costs across 6 LLMs and both baselines, normalized to \emph{Expert + SkyPilot}, to understand whether the approach produces consistent results across models. The four coding-specialized models, Opus4.6, Sonnet4.6, Qwen3.5, and Nemotron, are highlighted in Figure~\ref{fig:num1-geomean} as \emph{Coding LLMs}. The results reveal a clear stratification. The two commercial models achieve the lowest average instance costs, at 0.56$\times$ and 0.74$\times$ of the \emph{Expert + SkyPilot} baseline, respectively. Qwen3.5 also outperforms the expert baseline at 0.82$\times$, though with a higher job failure rate (see Figure~\ref{fig:num1-failures}). Nemotron ends up worse than the baseline at 1.18$\times$, despite being a coding-oriented model, suggesting that code specialization is necessary but not sufficient for this task.

Llama3.3 and Ministral3 are general purpose and produce geometric mean costs of 1.30$\times$ and 1.58$\times$ the \emph{Expert + SkyPilot} baseline, respectively. This tells us that model size alone does not guarantee recommendation quality. The 70B parameter Llama3.3 model is outperformed in all respects by the 35B parameter Qwen3.5 model, suggesting that code-specific training is more dispositive to outcomes than raw parameter count. \emph{SkyPilot} without expert constraints achieves a geometric mean cost of 1.13$\times$, outperforming the weaker LLM-based configurations but substantially underperforming the stronger coding models. The stronger coding models more consistently produce viable, cost-effective instance selections from submission-time artifacts alone, with a single prior run closing the residual gap. Across the board, commercial frontier models outperform open-weight alternatives, consistent with recent coding benchmark results~\cite{jain2025livecodebench, zhuo2025bigcodebench, qiu2024efficient}.

% ./plot_cost_num1.py

\section{Future Work}
\label{sec:future}

We currently target single-node job submissions using compiled binaries or Python scripts. Natural extensions include multi-node execution, requiring analysis of MPI communication patterns to recommend per-node instance types, node counts, and cloud-native interconnects; per-stage constraint estimation for DAG structured workflows with heterogeneous resource requirements; support for containerized applications by inspecting Docker or Singularity manifests, layer contents, and embedded binaries; and broader language facilities to support runtimes such as Julia, R or MATLAB. A separate direction concerns the scope of the recommendation itself. Incisor currently selects an instance type given a fixed invocation, yet opportunities may exist to co-optimize the execution environment and invocation alongside the hardware choice. Richer evidence sources such as source code, project documentation and web search could further sharpen constraint estimates. Exposing agent rationale to users and accepting natural language feedback would close a human-in-the-loop feedback path complementary to the automated loop already in place. Finally, supporting user-specified utility functions combining cost, runtime, and other objectives with performance modeling would allow Incisor to respond to richer expressions of user priority beyond high level cost and time reduction.

\section{Conclusion}
\label{sec:conclusion}

\noteh{Space Saving: Consider merging sections VIII and IX somehow?}

Incisor addresses the ex ante cloud instance selection via a 2-stage agent-driven solution, each grounded in tool use and independently verifiable by embedding them within an end-to-end system that handles provisioning, monitoring, and automatic fallback when selections prove unsuitable. Across 31 HPC workloads, frontier coding models achieve 100\% first-run success while reducing instance costs by 44\% and job runtime by 54\% relative to expert-derived constraints paired with SkyPilot, substantially outperforming general-purpose models. These results demonstrate that state-of-the-art agentic LLM systems can automate a task traditionally requiring scarce expert judgment, making cost-effective cloud HPC accessible to a broader range of users and workloads.

\section*{Acknowledgements}
Commercial and locally run agentic LLMs are a core component of our system and experimental infrastructure (Section~\ref{sec:methodology}). Claude, ChatGPT, and Perplexity Pro were also used as ``peer reviewers'' whose feedback was used to improve the manuscript prior to submission.

\bibliographystyle{IEEEtran}
\bibliography{main}

\end{document}